\documentclass[useAMS]{mn2e_modmargin}

\usepackage{amsmath}
\usepackage{url}
\usepackage{amsfonts}
\usepackage{amsbsy}
\usepackage{graphicx}
\usepackage{subfigure}
\usepackage{verbatim}
\usepackage{amssymb}

\newcommand{\Rr}{\ensuremath{R_c^\text{rad}}}
\newcommand{\Rt}{\ensuremath{R_c^\text{turb}}}

\renewcommand{\v}{\ensuremath{\mathbf{v}}}
\renewcommand{\b}{\ensuremath{\mathbf{b}}}

\newcommand{\B}{\ensuremath{\mathbf{B}}}
\newcommand{\FF}{\ensuremath{\mathbf{F}}}

\renewcommand{\u}{\ensuremath{\mathbf{u}}}
\renewcommand{\d}{\ensuremath{\partial}}

\newcommand{\ePhi}{\ensuremath{\mathbf{e}_{\phi}}}

\title[The dynamics of inner dead-zone boundaries
 in protoplanetary disks]
{The dynamics of inner dead-zone boundaries in protoplanetary disks}

\author[Henrik N. Latter]{Henrik N. Latter$^{1}$$^,$$^{2}$$^{\,}$\thanks{Email:
hl278@cam.ac.uk}, Steven Balbus$^{2}$\thanks{Email: steven.balbus@lra.ens.fr}\\ \\
$^{1}$ DAMTP, University of Cambridge, CMS Wilberforce Rd, Cambridge
CB3 0WA, UK \\
$^{2}$ LERMA-LRA, \'Ecole Normale Sup\'erieure, 24 rue Lhomond, 75231
Paris Cedex 05, France}

\begin{document}
 
\maketitle

\label{firstpage}

\begin{abstract}

 In protoplanetary disks, the inner radial boundary between the MRI
 turbulent (`active') and MRI quiescent (`dead') zones plays an important
 role in models of the disk evolution and in some planet formation
 scenarios.  In reality, this boundary is not well-defined: thermal heating
 from the star in a passive disk yields a transition radius close to
 the star ($<0.1$ au), whereas if the disk is already MRI active, it can
 self-consistently maintain the requisite temperatures out to a transition
 radius of roughly 1 au.  Moreover, the interface may not be static;
 it may be highly fluctuating or else unstable.  
 In this paper, we study a reduced model of
 the dynamics of the active/dead zone interface that mimics several
 important aspects of a real disk system. We find that MRI-transition
 fronts propagate inward (a `dead front' suppressing the MRI) if they
 are initially at the larger transition radius, or propagate
  outward (an `active front' igniting
 the MRI) if starting from the smaller transition radius.  In both cases,
 the front stalls at a well-defined intermediate radius, where it remains in a
 quasi-static equilibrium.  We propose that it is this new,
 intermediate
  stalling radius that functions as the true boundary between the
active and dead zones in protoplanetary disks. These dynamics are
likely implicated in observations of variable accretion, such as FU
Ori outbursts, as well as in those planet formation theories that
require the accumulation of solid material at the dead/active interface.

\end{abstract}

\begin{keywords}
  accretion, accretion discs --- instabilities --- MHD --- turbulence
  --- protoplanetary disks
\end{keywords}

\section{Introduction}

Protostellar disks, unlike the classic archetypes
 associated with dwarf novae and X-ray
binaries,
 are enormous, cold, and poorly ionised. In fact, significant
 regions within these disks are
so weakly coupled to the magnetic field that MRI turbulence
fails to develop (Balbus and Hawley 1991, Blaes and Balbus 1994, Gammie
1996, Igea and Glassgold 1999, Sano et al.~2000).
 Current models of protoplanetary disk structure consist of
a turbulent envelope of plasma (the `active zone')
 encasing an extensive body of quiescent
 gas (the `dead zone') (Gammie 1996, 
 Armitage 2011). In particular, these models posit a critical inner radius within
which the disk is fully turbulent, and beyond which the disk exhibits
a characteristic layered structure for some range of radii.

In general, it is unlikely that such a configuration supports
steady accretion onto the protostar (Gammie 1996, but see
also Terquem 2008). In fact, observations of YSOs show a rich
assortment of time-variable accretion, the most striking of which are the
outbursts associated with FU Ori and EX Lupi disks
(Hartmann and Kenyon 1996, Herbig 2008). In addition, measurements of
disk outflows reveal quasi-periodic variability on the order of
years to hundreds of years
 that may arise from fluctuating accretion in the inner disk,
precisely at those
radii where we might expect the inner dead/active zone boundary
(Lopez-Martin et al.~2003, Raga et al.~2011, Agra-Amboage et al.~2011).
        
Theoretical models that describe FU Ori behaviour attribute
an important role to the dynamics of this
 critical inner interface. In
particular, many models interpret
outburst events as the abrupt ignition of
the dead-zone in turbulence, and hence the temporary dissolution of the
critical boundary (Gammie 1996, Armitage et al.~2002,
Zhu et al.~2009, 2010). On the other hand, 
local numerical simulations of the MRI near
marginality --- the relevant regime at the interface ---
also exhibit strong variability in the form of violent
intermittency (channel flows) (Miller and Stone
2000, Sano and Inutsuka 2001, Latter et al.~2009, 2010) and oscillatory behaviour (Simon et al.~2011).
The simulations demonstrate that the MRI
will not smoothly `die away' as one crosses 
into the dead zone through the transition layer. Despite these
considerations,
the interface is assumed to be effectively \emph{static}
in a number of planet formation scenarios (Varni\'ere and Tagger
2006, Kretke et al.~2009, Dzyurkevich et
al.~2010). The fate of these theories is unclear if indeed the
interface 
is the site of
the regular and violent dynamics suggested by observations and simulations.

Our paper is concerned with the location, radial morphology, and dynamics
of this important disk feature. We concentrate on an
under-appreciated ambiguity in the location of the interface, one
that can potentially have consequences for the observations and
processes
 discussed
above. Gas at the midplane is thermally ionised 
by the central star
to sufficient levels for the MRI to
operate, but only out to a critical radius $\Rr$ very close to the star
($<0.1$ au). 
Beyond this radius lies a broader region in which stellar radiation fails
but ionisation by turbulent heating succeeds. However,
this requires there to be enough ionisation in the first place 
to get the
MRI turbulence initiated. If the gas in this region
 starts off cold and
poorly ionised it will remain so; conversely, if it begins hot and turbulent
it can sustain this turbulent state via its own waste heat. In other words, the gas
is `bistable': it can fall into one of two quasi-steady
states, active or dead. Finally, beyond a second critical radius $\Rt$
($\sim 1$ au)
neither stellar nor turbulent heating is sufficient and the disk
requires non-thermal sources to ionise the midplane. Usually this second
critical radius is taken to be the actual inner edge of the dead
zone. This need not be the case. The interface could in principle
fall within the bistable region at smaller radii, and it need not be
well-defined nor time-steady.

In order to study this physics we develop a reduced MHD model that
describes the interpenetration of the turbulence and the
thermodynamics. It generalises the evolution equations of Lesaffre et
al.~(2009), which couple the thermal energy with the turbulent energy,
and installs a spatial (radial) degree of freedom. The gross disk dynamics
are hence represented by two coupled reaction-diffusion equations. The
main prediction of this model is that the dead/active
interface, though well-defined, is not always stationary. It tends to travel
radially inward or outward depending on conditions at its present radial location.
Ultimately the interface comes to a halt at a special radius, some
65\% of the second critical radius $\Rt$. Moreover the interface is
relatively broad, with the temperature transition layer extending over
roughly 1 au. Numerical simulations should be able to check these
predictions, 
while also assessing the prevalence and importance of turbulent intermittency.

The organisation of this paper as follows.  In Section 2, we develop the
concept that gas in the inner zone of protostellar disks is bistable.  In
Section 3, we present a reduced model that displays the fundamental
behaviour of the dynamical system. This behaviour is more fully
elucidated in Sections 4 and 5, in which the model is explored and its
predictions presented. Next, the theory is applied, in
Section 6, to what we argue is a more realistic model for the inner region
of protostellar disks. Finally, in Section 7 we summarize our findings and
draw our conclusions.

\section{The inner regions of a protoplanetary disk:
 a bistable system}

This section develops the idea that gas in the inner disk can
fall into one of two states: a hot turbulent state, and a cooler
quiescent state. We obtain a criterion for the instigation of
MRI turbulence and show at which radii in the disk this is satisfied if we
consider (a) radiation from the central star, in the absence of
turbulence,
 and (b) thermal ionisation from turbulent dissipation, in the absence
 of stellar radiation. The latter radius can be significantly greater
 than the former, opening up a liminal region whose properties
 we discuss. For clarity, we focus on gas at the midplane only, but our
 calculations can be generalised to include the vertical structure of
 the disk.

\subsection{Criterion for the onset of MRI}

The MRI only functions when there is adequate coupling between the
gas and the magnetic field. An estimate of the critical ionisation
fraction necessary for this coupling can be obtained within the bounds
of resistive MHD. Hall effects are of comparable importance in certain
regions of the disk (Wardle 1999, Balbus and Terquem 2001, Wardle
and Salmeron 2012), 
but for the sake of clarity they will be neglected; we expect no
qualitative change in our calculation.

The shortest MRI-active mode works on the
resistive scale: $l_\eta= \eta/v_A$, where $\eta$ is the Ohmic resistivity and
$v_A$ is the Alfv\'en speed. This lengthscale cannot be larger than the disk
scale height $H$. So we have marginal MRI when $l_\eta=H$ which gives 
\begin{equation}
\mathcal{L}\equiv v_A H/\eta = 1,
\end{equation}
where $\mathcal{L}$ is the Lundquist number. To turn this into a critical
ionisation fraction we take $H\approx 0.1 R$ and the following expression for
resistivity
\begin{equation}\label{etadef}
 \eta = 234\, x_e^{-1} T^{1/2}\, \text{cm}^2\,\text{s}^{-1}, 
\end{equation}
where $x_e$ is electron fraction and $T$ is temperature (Blaes and Balbus 1994). Combining
these equations gives the critical ionisation level:
\begin{equation} \label{xe}
(x_e)_\text{crit} = 1.89\times 10^{-14}\,\beta^{1/2}\,R_\text{AU}^{-1},
\end{equation}
where $\beta$ is the plasma beta and $R_\text{AU}$ is disk radius in units of au (see also
Balbus 2011). At 1 au and
$\beta=10^2-10^3$ we have the fiducial limit of $x_e\sim 10^{-13}$. 

Ionisation in a protoplanetary disk can be caused by non-thermal sources,
such as 
cosmic rays, stellar X-rays, radionuclides, and energetic stellar
protons (Stepinski 1992, Gammie 1996, Igea and
Glassgold 1999, Turner and Drake 2009), in addition to direct heating
from either the central star or turbulent dissipation.
 Each source gives rise to a spatial ionisation structure in
the disk. In the case of thermal ionisation, this structure 
follows directly from
the disk's temperature profile, and hence
we can  associate a critical temperature $T_\text{MRI}$
 with the critical ionisation
fraction $(x_e)_\text{crit}$.
The critical temperature may be computed with the aid
of an appropriate form of Saha's equation, noting that the ionisation
is controlled by the low ionisation-potential alkali metals (Pneuman
and Mitchell 1965, Umebayashi and Nakano 1988). A straightforward
calculation (e.g.\ Stone et al.~2000 or Balbus 2011)
 shows that Eq.~\eqref{xe} translates into
\begin{equation}\label{T_MRI}
T_\text{MRI} = 800\, -\,1000\, \text{K}.
\end{equation}
Throughout the paper we take $T_\text{MRI}=900$ K as our reference.

\subsection{Ionisation from turbulent heating}

MRI turbulence, if instigated, can
 thermally ionise the gas with some of
its waste heat. The MRI can thus
positively reinforce the unstable state from which it springs. In the
complete absence of radiation, the MRI produces enough heat to keep the
gas sufficiently ionised out to the critical radius
$R_c^\text{turb}$ (Kretke et al.~2009, Balbus 2011). Essentially, the
turbulence thermalises the free energy locked up in
 the background orbital shear.
 However, the available free energy declines with radius, and
beyond $\Rt$ there is not enough
 to keep things sufficiently hot if thermalised (given
realistic disk opacities).

The critical radius $\Rt$ can be calculated via an alpha disk model,
which supplies a monotonically decreasing 
radial temperature structure for the disk. The radius
at which point $T=T_\text{MRI}$ corresponds to $\Rt$. 
Of course, beyond $\Rt$ the alpha model is no longer
consistent, because accretion via the MRI no longer functions in this
region at every vertical level, but as a first approximation the approach is adequate.
Kretke et al.~(2009)
summarised their numerical results by the following analytic estimate:
\begin{equation} 
\Rt \approx
0.52\,\dot{M}_{-7}^{4/9}\,M_*^{1/3}\,\left(\frac{\alpha}{10^{-2}}\right)^{-1/5}
\,\left(\frac{\kappa}{\text{cm}^2\,\text{g}^{-1}}\right)^{1/4} \,\text{au},
\end{equation}
where $\dot{M}_{-7}$ is the mass accretion rate scaled by $10^{-7}$ solar masses per year,
the stellar mass $M_*$ is scaled by solar mass, and $\kappa$ is the
Rosseland mean
opacity.
 With the positive
correlation between $\dot{M}$ and $M_*$, the more massive the star
 the greater $\Rt$ (Kretke et al.~2009). 
In summary, the critical radius associated with MRI turbulence falls
between about 0.1 and 3 au.

\subsection{Ionisation from stellar radiation}

Consider now a passive disk, in which there is no turbulence or
accretion. It is irradiated by the central star
 and, potentially, cosmic rays.
Though non-thermal sources, such as stellar X-rays and cosmic rays,
 are dominant in the surface layers of such a
disk at small radii (and throughout the disk at large radii),
 midplane
gas at $R<1$ au is well shielded and ionisation
rates are vanishingly small.
As a consequence, the associated $x_e$ is tiny (given standard dust-grain chemistry)
 and the MRI is unable to work 
(Igea and Glassgold 1999, Sano et al.~2000, Ilgner and Nelson 2006, Wardle 2007,
 Turner and Sano 2008,
 Turner and Drake 2009). More important to midplane gas at these radii
  is stellar thermal radiation.

A number of sophisticated radiative transfer studies have
calculated the temperature structure of
a passive disk irradiated by a realistic stellar source (see Pinte et
al.~2009 for references). These calculations omit turbulent transport
in the active surface layers, but in any case
 the simulations of Hirose and Turner (2011) show that
vertical turbulent transport of heat is negligible. The 
 disk is usually truncated at a defined inner radius and the midplane
 temperature calculated as a function of distance from this inner
 edge. As discussed by Dullemond and Monnier (2010), at first the temperature
 structure is
 controlled by the dust sublimation threshold. However, once the
 sublimation temperature is reached, and dust can
 survive, $T$ drops rapidly with radius.  
The temperature profiles calculated by Pinte et al.~(2009)
  suggest that
  the midplane gas falls below
 $T_\text{MRI}$ when it is more than 0.01 au from the inner edge.
This result is remarkably robust across a range of 
normal optical depths ($10-10^6$) and different extant codes. This
finding is reinforced by the more
detailed calculation of Woitke et al.~(2009) which includes a
self-consistent account of the disk vertical structure and a great
deal more radiative physics. These calculations imply that $\Rr$ may plausibly 
lie between $0.01$ and $0.1$ au, if the disk indeed has an inner edge. But if the disk
physically connects to the central star, then $\Rr$ may be
less. Given these uncertainties, we set a (crude) upper bound on
$\Rr$ to be 0.1 au.

\subsection{The bistable region}

The ordering $\Rr < \Rt$ invites us to split the inner region
of a protostellar disk into three zones as sketched out in Fig.~1.
 At very small radii $R<\Rr$, the disk will always
be sufficiently irradiated by the star to ionisation levels that
sustain the MRI. Similarly, the surface layers will also be unstable.
 Heat dissipated from turbulence is not required in these regions. At larger radii
$R>\Rt$, however, the miplane gas will always be too cold and poorly ionised to
sustain the MRI. On one hand, it is adequately shielded from the star's
radiation field. On the other, even if we managed somehow to
kickstart the MRI at these radii, it
would eventually die out because its waste heat is insufficient to
sustain the necessary temperatures.

\begin{figure}
\scalebox{0.5}{\includegraphics{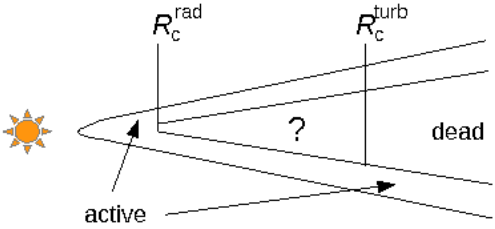}}
 \caption{Cartoon of a three zone disk. The very inner radii
   $R<\Rr\approx 0.1$ au (and also
   surface layers) are
   sufficiently ionised by the star to always support MRI. The outer
   radii $R> \Rt \approx 1$ au are always too cold to support the MRI --- the turbulence cannot heat
   the gas sufficiently in this region and it is well shielded from the
   star. The middle region, sandwiched between the other two could go either
   way --- turbulent or quiescent. It is `bistable'. Note that at
   sufficiently large radii $R\gtrsim 10$ au  (not included in the
   cartoon)
  the disk may again be fully MRI unstable.}
\end{figure}

This picture leaves an ambivalent region $\Rr<R<\Rt$ sandwiched
between the other two. Here the gas can be either quiescent or
turbulent. It is \emph{bistable}. If the gas starts off cold and
quiescent \emph{it will remain cold and quiescent}: the star's radiation cannot instigate the MRI in
this region on its own. But if the gas starts off
 hot and turbulent \emph{it will
remain hot and turbulent}: the MRI will dissipate heat and this heat will be
sufficient to maintain temperatures above the critical limit. The gas
can only transfer from one state to the other via a sufficiently large (nonlinear)
perturbation.

This is a non-trivial point with potentially important ramifications.
It leaves ambiguous the actual location of the dead/active zone
boundary, and moreover implies that there may not
be a well-defined interface at all.  For instance, 
the gas here may oscillate
between the two limits, or exhibit a range of interesting
dynamical phenomenon, by analogy with other bistable systems
(Zeldovich and McNeill~1985,
Murray 2002). More generally, nonsteady behaviour in this region could explain some of
the low amplitude accretion variability observed, and will probably 
 be implicated in the dynamics of outburst events, if they are indeed
 caused by the aperiodic dissolution of the dead-zone (Gammie 1996).

\section{A reduced model for the turbulent/thermal dynamics}

Our goal is to describe the dynamics in the bistable region of the
disk, in particular 
the interplay between the
thermal/ionisation dynamics and the turbulent dynamics.
To that end we construct a mean-field model that captures this
qualitative behaviour correctly. 
 Our reduced model is rather crude, especially in its treatment of
 the turbulence,
yet it
illustrates in a clean way some basic ideas that should
govern more advanced models and simulations.

 We the define the turbulent amplitude of the disk gas to be
\begin{equation}
K= \tfrac{1}{2}\langle  \rho \v^2 + \b^2/(8\pi)  \rangle,
\end{equation}
where $\v$ and $\b$ are fluctuations in velocity and magnetic field,
$\rho$ is density,
and the angle brackets denote an azimuthal and vertical average over
the disk, as well as an average over short radial scales (see
Appendix A).
Both $K$ and the gas temperature $T$ are intertwined in the bistable zone: the magnitude of the turbulent amplitude
will determine the temperature via turbulent dissipation, but the
magnitude of the temperature will determine the ionisation fraction
and consequently whether turbulence is present
or not. These relationships make up the `reaction terms' in our
dynamical model, and have been investigated on their own previously by
 Balbus
and Lesaffre (2008) and Lesaffre et al.~(2009). In addition, we add the influence of \emph{turbulent diffusion},
treating $T$ and $K$ as reactive scalars in a turbulent flow field which
both transports and is reacted upon by $T$ and $K$.

 In effect, we
construct a turbulent closure model with a novel dynamical connection to
the thermodynamics. Analogous models have
been used in various terrestrial contexts with some success, such as the $k-\epsilon$
turbulence model (for example, Davidson 2000), the
multi-scale fluid models (see Frisch 1995 for references), and in
mean-field plasma turbulence (for example, Gruzinov and Diamond 1994). Of
course, the closest cousin of the model we present is the alpha disk
itself, as interpreted by Balbus and Papaloizou (1999), and its more
complicated variants (Armitage et al.~2002, Wunsch et al.~2005, 2006,
 Zhu et al.~2009, 2010). 
The detailed derivation of our governing equations we give in Appendix A; in the
main body of the paper  we
motivate their main features heuristically. We begin their exposition
with a summary of the closely related system in Lesaffre et al.~(2009).

\subsection{Homogeneous dynamics}

 Lesaffre et
al.~(2009) argue that the full equations of viscous and resistive MHD in a
shearing box can be effectively modelled by
 the following simple dynamical system:
\begin{align} \label{hom1}
\frac{d K}{d t} & = Q(K,T),  \\
\frac{d T}{d t} & = \Gamma(K) - \Lambda(T), \label{hom2}
\end{align}
where $K$ is the box-averaged turbulent kinetic and magnetic
energy,
 and $T$ is the box-averaged temperature. The evolution of
these quantities is controlled by three functions: $Q$, which
summarises the growth and saturation of the MRI, $\Gamma$ which
represents turbulent heating, and $\Lambda$ which represents radiative
cooling. Though these functions can be written as complicated averages
over the fluctuating quantities, they may be more simply
 modelled by physically motivated algebraic expressions. When
quantitatively compared to full MHD simulations, this simple system
does remarkably well in describing the principle dynamics (Lesaffre et
al. 2009), which inspires confidence in its utility. 

\subsubsection{Parametrising the MRI }

To describe the onset and
saturation of the MRI via the term $Q$
a simple Landau operator is invoked for a single mode amplitude, which yields
\begin{equation}
 Q \approx s\,K - a\,K^2,
\end{equation}
where $s$ is the MRI linear growth rate and $a$ is a parameter associated
with its saturation.  The temperature influences the linear forcing
term, the growth rate, via
\begin{equation}
 s = s_0[1- \overline{\eta}(T)],
\end{equation}
where $s_0$ is the ideal MHD growth rate, and $\overline{\eta}$ is the
magnetic diffusivity scaled by a constant reference diffusivity.
Lesaffre et al.\ use the Spitzer prescription for $\eta$ but in
weakly ionised protoplanetary disks one based on an appropriate Saha equation is
required (outlined in Appendix A).

\subsubsection{Turbulent heating}

 The turbulence taps energy stored in the background
differential rotation. Turbulent stresses
 can transform this energy into fluctuations that tumble
down a cascade (or pseudo-cascade) into the arms of the dissipation
scales where it is degraded into heat. The rate of energy injection
into the system is hence proportional to the quadratic correlation
$\langle v_R v_\phi- b_{R} b_{\phi}/(4\pi\rho) \rangle $.
Lesaffre et al.\ approximate this correlation by the turbulent magnitude $K$ (itself a
quadratic quantity in the fluctuations) and write 
\begin{equation}
\Gamma = w\,K,
\end{equation}
where $w$ measures the local shear rate. The
greater the local shear the more energy the turbulence can extract. We
hence expect $w$ to be a decreasing function of radius, though in most
applications we will leave it as a constant.

\subsubsection{Radiative cooling}

Finally, if we assume that the disk surfaces radiate like a blackbody,
 the cooling function $\Lambda$ may take the following profile
\begin{equation}
\Lambda = b\,(T^4 - T^4_\text{eq} ),
\end{equation}
where $T_\text{eq}$ is the temperature at radiative equilibrium and $b$ is
some free parameter associated with the opacity of the gas. It
controls, in basic terms, the speed at which radiative equilibrium is
enforced. The cooling time is hence $\sim 1/(b T^3)$, which we expect
to be generally much longer than an orbit in the bistable region.
 A small optical thickness corresponds to a
large $b$ and a large optical thickness
corresponds to a small $b$. Consequently, $b$ may be
 a function of $R$ in general. Note
that in Lesaffre et al. (2009) a simpler linear cooling was employed.

\subsection{Turbulent transport}

Equations \eqref{hom1}-\eqref{hom2}
 describe the average
dynamics at a fixed radius in the disk, however our goal is to capture
the disk evolution over a significant range of radii and when subject to turbulent
transport. Certainly, MRI turbulence effectively moves angular
momentum outward, but it will also transport heat, though
perhaps less efficiently.  The MRI will also tend to spread kinetic
and magnetic energy: the more aggressive
 turbulent fluctuations in higher intensity regions will travel into
less turbulently intense regions. The latter transport is analogous to
the spreading of kinetic energy away from a region of localised
forcing and into unforced quiescent gas. The disordered motions decay as
they spread, but these residual motions still 
have energy distributed over a range of scales which ultimately dissipates.
 
The simplest way to model the
transport of $T$ and $K$ is as a diffusive process. We may then
 permit $T$ and $K$ to vary with radius $R$, in addition to
time, and subsequently erect Fickian diffusion (or `eddy viscosity')
 terms in both equations
\eqref{hom1} and \eqref{hom2}. This augmented model would then resemble
\begin{align}
\frac{\d K}{\d t} & = Q + \frac{1}{R}\frac{\d}{\d R}\left( R\,D_K\,\frac{\d K}{\d
    R} \right), \\
\frac{\d T}{\d t} & = \Gamma - \Lambda + \frac{1}{R}\frac{\d}{\d R}\left( R\,D_T\,\frac{\d T}{\d
    R} \right),
\end{align}
in which we have introduced diffusion coefficients $D_K$ and
$D_T$. Note that the latter coefficient not only incorporates
transport by turbulent eddies but also by
radiation. Both $D_K$ and $D_T$ should be (increasing) functions of the turbulent
intensity $K$, and $D_T$ may depend on temperature as well. However, for
simplicity, we will assume that they are constant.

\subsection{Dimensionless equations}

The number of free parameters in the system can be reduced by choosing
suitable dimensions. We scale time by $1/s_0$ which is of order an orbit, and we scale
turbulent amplitude $K$ and temperature $T$ by reference values $K_\text{eq}$
and $T_\text{eq}$. We denote by $K_\text{eq}$ the turbulent amplitude in the
absence of resistivity (recall that $T_\text{eq}$ is
 the temperature in the absence of turbulence). The quantities $s_0$,
 $T_\text{eq}$, and to some extent $K_\text{eq}$, may be associated with a
prescribed disk radius lying within the bistable region, between 0.1 and 1 au. 
 Next we set the diffusivities 
$$ D_T = L_T^2\,s_0\, \qquad D_K= L_K^2\,s_0
$$
where $L_K$ and
$L_T$ are the associated diffusion lengths at the prescribed radius. We next scale the unit of
length in our equations by $L_T$. Together these transformations yield the simpler set
of reaction-diffusion equations:
\begin{align}
\frac{\d K}{\d t} &= \overline{s}K - K^2 + P_r\,\frac{1}{R}\frac{\d}{\d
  R}\left(R\,\frac{\d K}{\d R}\right), \label{finK} \\
\frac{\d T}{\d t} &= \overline{w}\,K - \overline{b}(T^4-1) + \frac{1}{R}\frac{\d}{\d
  R}\left(R\,\frac{\d T}{\d R}\right). \label{finT}
\end{align}
Here 
$$ \overline{w}= \frac{w\, K_\text{eq}}{s_0\,T_\text{eq}}, \qquad \overline{b} =
\frac{b T_\text{eq}^3}{s_0}, $$
and we have defined the Prandtl type number $P_r = L_K^2/L_T^2 $. From
Appendix A, we have a growth rate prescription approximating the
influence of the
Saha ionisation law
\begin{equation}\label{grorate} 
 \overline{s} =
 \text{tanh}\left[6(T-T_\text{MRI})\right].
\end{equation}
This means that near the temperature $T_\text{MRI}$ the ionisation
fraction jumps rapidly to a value sufficient to instigate the MRI. 
If, in physical units, $T_\text{eq}\approx 250$K (associated perhaps with
$R=0.5$ au), we have
$T_\text{MRI}\approx 3.6$. 
There are now three parameters $\overline{w}$, $\overline{b}$, and
$P_r$. For notational ease
 the overlines over $\overline{s}$, $\overline{w}$, and
$\overline{b}$ will be suppressed in the following.
Finally, we adopt $P_r=1$, not only
for simplicity but also because we have no
 numerical estimates for the relative efficiency of
turbulent $K$ and $T$ mixing. Thus $L_T=L_K$. 
The model offers no detailed quantitative constraints on
the two remaining parameters $w$ and $b$, but their meanings are
relatively transparent, measuring shear rate and inverse 
optical thickness, respectively. Because the injection of energy via
turbulence proceeds on a time scale longer than an orbit, we take
$w<1$. Similarly, the cooling time $1/(b T^3)$ must be longer than the
orbital time. With $T=T_\text{MRI}$, this yields $b< 0.01$.

\subsection{A further simplification: slavery}

It is possible to reduce the system \eqref{finK}-\eqref{finT} even
further if we assume that (a) the time-scale of the MRI saturation is much
faster than the thermal timescale, and (b) that the diffusion of $K$
 is slow compared to the diffusion of $T$, i.e. $P_r\ll 1$. 
Neither assumption may be
 strictly true, but the simpler system is particularly convenient to
 analyse, allowing us an insight into how the more
 general system works.

These assumptions mean that $K$ is always in equilibrium
 on the long thermal times that concern us. In other words the turbulent $K$ dynamics are `slaved'
to the thermal $T$ dynamics. For long times we have
\begin{equation}
K = \begin{cases}
   0 & \text{if}\quad T \leq T_\text{MRI} \\
   s & \text{if}\quad T> T_\text{MRI} \,.
\end{cases}
\end{equation}
Now we need only solve \emph{one} evolution equation, which we
write as
\begin{equation} \label{slavery}
\frac{\d T}{\d t} = \Gamma - \Lambda + \frac{1}{R}\frac{\d}{\d
  R}\left(R\,\frac{\d T}{\d R}\right),
\end{equation}
where as before the cooling rate is $\Lambda = b(T^4-1)$ and now the
heating rate is the continuous piecewise function:
\begin{equation}
\Gamma = \begin{cases}
   0 & \text{if}\quad T\leq T_\text{MRI}  \\
   w\, s & \text{if}\quad T> T_\text{MRI} \,.
\end{cases}
\end{equation}

\subsection{Caveats}

Before applying the model we should stress a few points regarding
 some of its weaknesses, especially with respect to its simplistic
 treatment of the turbulence.
As is made clear in Appendix A, the validity of the model rests on
the soundness of averaging over the small-scale disordered motions 
of the MRI turbulence, in addition to the vertical thickness of the disk. 
In particular, it demands a clean separation between large
scales of at least $H$, upon which global disk properties manifest,
 and the short MRI turbulent scales.
 Stratified simulations of the MRI
suggest this is only marginally true at best (Davis et al.~2010, for
example), while near criticality the fastest growing mode varies on approximately
$H$.
 We believe this problem will not derail the
qualitative results.

In addition, for the small scale
averages to make sense, 
the fluctuations need to be well-behaved ---
they cannot be subject to intermittent and volatile outbursts or other
critical behaviour. Yet that is what we might expect from the
marginal MRI! The best we can hope for is that this bad behaviour does
not invalidate the predictions of the mean-field model --- that, in
fact, the model indeed does a decent job of describing the mean
dynamics, even if the fluctuations around that mean are a little
wild at times. 

One should also add that
 the eddy viscosity description of turbulent flow has long been
known to be at best a crude, and sometimes misleading, approximation.
 In reality the stress exerted by
such flows are non-Newtonian, sometimes exhibiting a nontrivial
dependence on the shear rate, effects associated
with their finite relaxation time, and even negative transport (see,
for example, Frisch 1995, Davidson 2000). That all said, we are
interested 
only in the basic
dynamics, which we believe are independent of the
particulars of the turbulence itself.

 A final issue is the vertical averaging: in doing this,
 we are discarding any information about how the disk
thickness responds to local changes in either the turbulence or
temperature. It is implicitly assumed that $H$ varies
passively with
changes in $T$. For example, a transition between the hot MRI active
region and a cold quiescent region will be characterised by a
drop in $H$, meaning the inner turbulent radii of a
protostellar disk may bulge. Any
interesting stability problems, or other dynamics, that might arise
from this configuration
 cannot be captured by
the model.

\section{Analysis of homogeneous states}

\subsection{Local states }

Our first task is to
compute \emph{homogeneous} steady states --- i.e.\
equilibrium states with no
radial structure, which describe either dead or active zones.
As in a shearing box, such equilibria we expect to hold locally in a
protoplanetary disk.
 Because the diffusion terms take no part in these
calculations, we need only solve a coupled pair of ODEs. 

The
steady-state equations to be solved are:
\begin{align}
 s K - K^2 =0, \qquad
 w\,K - b(T^4-1)=0,
\end{align}
which arise directly from Eqs \eqref{finK}-\eqref{finT}.
 The first equation
matches the linear forcing by the MRI to its saturation through the
quadratic term. The second equation matches the first turbulent heating
term with the second radiative cooling term.

 This system admits
 the cold quiescent (dead) state
$$ K=0, \qquad T=1. $$
Turbulence is absent and the gas falls into radiative equilibrium.
The system also
potentially admits turbulent
states $K>0$, with $K=(b/w)(T^4-1) $ and with the temperature determined from
the nonlinear algebraic equation 
\begin{equation}\label{bal}
 w\,s = b(T^4-1),
\end{equation}
with $s$ given by Eq.~\eqref{grorate}.
Here the heating rate $\Gamma=w\,s$ is on the left and the cooling rate
$\Lambda=b(T^4-1)$ on the
right. Naturally, the linear growth rate $s$ must be positive for
this to work or else there is no turbulent heating at all. Note also
that the equilibrium depends on a single parameter, the ratio $b/w$.

Equation \eqref{bal} must be solved numerically. But if we plot the heating and
cooling as functions of $T$ we can get a good idea of how the balances
work. Examples are given in Fig.~2 for three representative values of $b/w$.
The points at which the two curves cross correspond to
turbulent/thermal equilibria.
Note that, depending on the ratio $b/w$, we can achieve one solution
(the cold quiescent state), two solutions, or three solutions. In the
figure we plot these three different cases. 
 The critical $b/w$ value above which turbulence
vanishes, and for which there is only the dead state, is approximately
$0.0042$. When $b/w>0.0042$ the cooling rate is too efficient,
  and/or the amount of free energy insufficient, for the requisite MRI
temperatures to be maintained.
Naturally
we associate the cold quiescent state with a dead zone, and the hotter
of the two turbulent state with the active zone. The intermediate state we show
later is unstable and will not be observed, though it is important in
policing the boundary between the other equilibria.

\begin{figure}
\scalebox{0.5}{\includegraphics{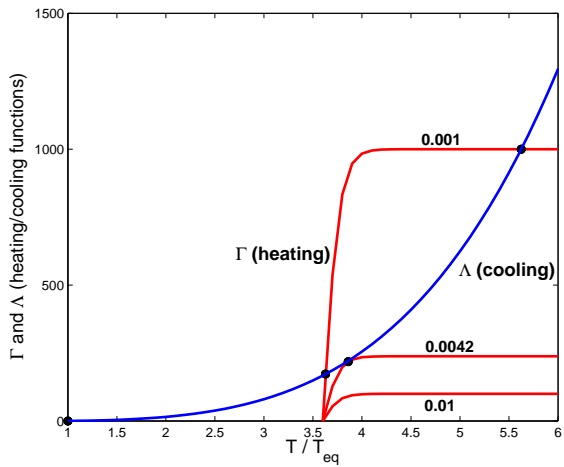}}
\begin{footnotesize}
 \caption{Heating and cooling as functions of
     temperature $T$ for different ratios of $b/w= 0.001,\,0.0042,
     0.01$. 
   Crossings correspond to thermal equilibria and
     are marked with black dots. These comprise the cold quiescent state
   $T=1$, and up to two turbulent states: $T_I$ (which we term
   the `intermediate state') and $T_A$ (which we call the `active state'). 
 The former is unstable and the latter is stable.}
\end{footnotesize} 
\end{figure}

Another way to characterise the equilibria is to rewrite Eq.~\eqref{bal} as
$$ T= \left[1+ (w/b)s\right]^{1/4}.$$
Because $s$ is essentially a step-function in $T$, and the upper
turbulent solution always occurs when resistivity has been virtually
`switched-off', we can approximate the upper state $T_A$ by
\begin{equation} \label{EqmApprox}
T_A \approx (1+w/b)^{1/4}, \qquad K_A \approx 1.
\end{equation}
Therefore, the greater the parameter combination $w/b$ the greater the
active zone temperature $T_A$.
 This makes sense because $w/b$
 measures the amount of heat stored in the gas. 
 Recall $b$ represents the ability of the gas to hold heat
 (optically thick gas takes smaller $b$) while $w$ controls the
 amount of heat generated via turbulence ($w$ is larger when more energy is available to
be dissipated). 
Optically thicker gas at smaller radii is hotter, whereas optically thinner gas at larger
radii is colder.

\subsection{Linear stability}

Next we briefly inspect the stability of the three possible
steady states. We find that the quiescent state $K_Q=0$,
$T_Q=1$ is always stable on account of its cooling law (heating is
absent around this state). The upper turbulent state $K_A\approx
1$, $T_A\approx (1+ w/b)^{1/4}$ is also stable. This is easy to see from
the heating and cooling functions in Fig.~2. If the temperature
increases from the hotter active state the cooling function will
be greater than the heating function: the excess temperature
will be removed and the system will return to equilibrium. On the other hand, 
if the temperature decreases then heating will dominate
cooling and the gas will also return to equilibrium. 
 It is quite the
opposite
for the intermediate equilibrium: a small positive deviation in temperature means
that heating will dominate cooling and so the gas will continue
heating up. Similarly a small negative deviation means that cooling
will dominate.

This heuristic reasoning can be confirmed with a detailed linear
stability analysis, which we present in Appendix B. 
 The linear dispersion relation that ensues gives the
following
 stability criterion
\begin{equation} \label{stability}
\frac{d\Lambda}{d T} > \frac{d\Gamma}{d T},
\end{equation}
which puts in mathematical terms the graphical argument presented
earlier. Direct substitution of $\Gamma= w K$ and
$\Lambda= b(T^4-1)$ shows that the quiescent state and the upper
active state are always stable, while the intermediate state is not.
 It is, in fact, a \emph{saddle point} with one stable and one
unstable mode (manifold).

\subsection{Nonlinear homogeneous dynamics}

Lastly, we describe the nonlinear dynamics of the homogeneous
equations. This
situation might mimic
conditions in a localised patch, as approximated by a shearing box
(cf.\ Lesaffre et al.~2009). 
The full dynamical system here is:
\begin{align}
\frac{d K}{d t} &= s\,K - K^2, \label{ph1}\\
\frac{d T}{d t} &=  w\,K - b(T^4-1), \label{ph2} 
\end{align}
two nonlinear ODEs in time, which we solve numerically using a
Runge-Kutta method. In Fig.~3 a number of numerical solutions to this
system are plotted
alongside the vector `reaction' field $\mathbf{R}$ which controls the solution
trajectories:
\begin{equation}\label{RRR}
\mathbf{R} = [s\,K - K^2,\, w\,K - b(T^4-1)].
\end{equation}
The three equilibrium points are represented by black spots.

\begin{figure}
\scalebox{0.45}{\includegraphics{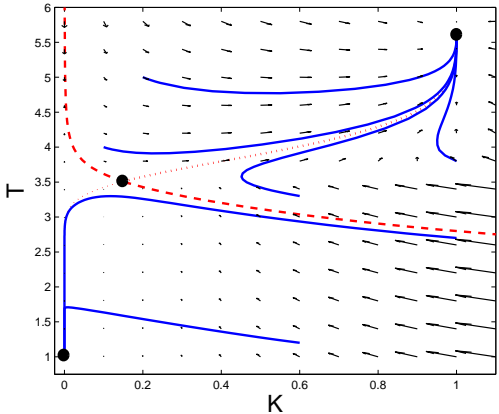}}
\begin{footnotesize}
 \caption{Phase portrait of the homogeneous problem. Some solutions of
 Eqs \eqref{ph1} and \eqref{ph2} are plotted in blue along with the reaction
 vector field in black. The three equilibrium solutions are represented
 by black dots. In addition,
 the stable and unstable manifolds of the intermediate state are
 plotted as dashed and dotted lines respectively. Trajectories which
 begin above the dashed line (the stable manifold) end up at the
 turbulent state, and trajectories which begin below the dashed line
 end up in the dead state. The average time it takes to reach either
 state is limited by the cooling time. A trajectory settles on the hot
 turbulent state in some 10 orbits, whereas one approaches the
 quiescent state in over 100 orbits. 
Parameters are $w=1$, $b=0.001$.}
\end{footnotesize} 
\end{figure}

As is clear, trajectories eventually end up on one of the stable
states. The speed at which
the system moves to one or the other state is limited by the cooling
rate $\sim b T^3$.
 Most trajectories are first attracted to the unstable saddle
point, guided along its stable manifold (the red dashed curve),
 but are finally repelled along the
stable manifold (the dotted curve). 
Which state they terminate upon depends on where in the phase
space they begin and, more specifically, in which basin of attraction
they initially fall.

 The phase space can be divided into two regions,
the basin of attraction of the quiescent dead state, and that of the
active turbulent state. The boundary between the two is given by the
stable manifold of the intermediate state (the dashed curve).
As the parameters change, this boundary also changes, with the basins
of attraction growing or shrinking relatively. In particular, as $b/w$
decreases the intermediate state approaches the low state, thus
diminishing the basin of attraction of the low state.

\section{MRI-fronts}

We have dealt with the homogeneous
dynamics of our reduced model, first detailing the equilibrium states
the system admits, their linear stability, and the ensuing nonlinear
dynamics. The analysis has been limited
to gas at a particular radius, allowing the gas to settle on
either an active or dead state. But, as we vary the radius
in the disk, the gas presumably transitions from one state to
the other at some sort of interface. 
In this section we attempt to compute
`fronts' connecting these two stable homogeneous
equilibria. We hence must reinstate the spatial element of the
nonlinear problem. 

\subsection{Equations in a semi-local model}

In order to best study a nonlinear front solutions
we concentrate on a small finite region at a certain radius $R_0$ in the
disk. We assume that any kind of front structure will be
much shorter than the local radius. We hence drop the term
arising from the 
cylindrical geometry in Eqs \eqref{finK}-\eqref{finT}.
We also introduce an intermediate radial variable $X$ so that
$$ X= R-R_0.$$
The equations are now
\begin{align}
\frac{\d K}{\d t} &= s\,K - K^2 + \frac{\d^2 K}{\d X^2}, \label{frK} \\
\frac{\d T}{\d t} &= w\,K - b(T^4-1) + \frac{\d^2 T}{\d X^2}. \label{frT}
\end{align}
Next we make the following ansatz: there exists a front with a radial
profile that is steady and which 
moves radially at a constant speed
$c$. This front connects the stable quiescent equilibrium state
$(K_Q,\,T_Q)$ and the stable turbulent state $(K_A,\,T_A)$. 
Because of radial disk structure, the assumptions of constant $c$ and
steady profiles
  only hold approximately in our small region
near $R_0$. However, they permit us to directly compute the fronts and
to clearly examine their properties. We relax these assumptions 
later in numerical simulations.
 
To make progress we move into a frame comoving with the front, and 
introduce the comoving spatial variable
\begin{equation}
\xi = X - c\,t.
\end{equation}
In this frame the front profile is stationary and thus $K=K(\xi)$ and
$T=T(\xi)$. The partial differential equations \eqref{frK}--\eqref{frT}
become the coupled ordinary differential equations
\begin{align}
&\frac{d^2 K}{d\xi^2} +
c\frac{dK}{d\xi}+ s\,K - K^2 = 0, \label{frKK}\\
&\quad \frac{d^2 T}{d\xi^2} +c\frac{dT}{d\xi}
+ w\,K - b(T^4-1)=0.\label{frTT}
\end{align}
 The boundary conditions for
these equations come from the requirement that the front must connect
$(K_Q,\,T_Q)$ with $(K_A,\,T_A)$. Formally at least, this may be written
as 
\begin{align}
&K\to K_Q, \quad T\to T_Q \quad \text{as} \quad \xi\to\infty, \label{bc1}\\
&K\to K_A, \quad T\to T_A \quad \text{as} \quad \xi\to -\infty.\label{bc2}
\end{align}
A more compact vector form for the equations \eqref{frKK} and
\eqref{frTT} is
\begin{equation}\label{compaq} 
\frac{d^2 \mathbf{z}}{d\xi^2} +
c\frac{d\mathbf{z}}{d\xi}+ \mathbf{R} = 0
\end{equation}
where we haved introduced the solution vector $\mathbf{z}=[K,\,T]$.
 Recall that $\mathbf{R}$ is
the reaction field introduced in \eqref{RRR}. It corresponds to the
trajectory arrows in Fig.~3.
 
Equations \eqref{frKK}-\eqref{bc2} describe a
 one-dimensional nonlinear eigenvalue problem
for $K$ and $T$ with eigenvalue $c$. These equations must be solved
numerically. We give the results of such
computations in a following subsection. But first we prove some
 results concerning the direction of front propagation.

\subsection{Direction of front propagation}

The hypothetical front solution may not be static. But in
which direction will it move and at what speed? If it connects a quiescent state
to a turbulent state, will it move into the turbulent state $(c<0)$ 
or will it
move into the quiet state $(c>0)$? Later, it is shown numerically
 that either case is
possible. In this section we try and provide some arguments for why
this is so and offer some analytic predictions for what sign $c$
will take.

The main ideas
 can be best illustrated using the simple
`slaved' system introduced in Section 3.4.
Upon imposing the assumptions of the previous subsection,
 the governing equation \eqref{slavery} becomes
\begin{equation}\label{slappy}
\frac{d^2T}{d\xi^2}+ c\frac{dT}{d\xi}
+ \Gamma-\Lambda = 0,
\end{equation}
Multiplying
by $dT/dz$ and integrating, one gets
\begin{equation}\label{slav1}
c= -\frac{\int_{-\infty}^\infty (\Gamma-\Lambda)\d_\xi T\,d\xi}{\int_{-\infty}^{\infty}|\d_\xi\,T|^2 d\xi}.
\end{equation}
and so
\begin{equation}\label{slavec}
 \text{sgn}(c) = \text{sgn}\int_{T_Q}^{T_A}(\Gamma-\Lambda)\,dT.
\end{equation}
 This tells us that the sign
of $c$ depends on the `area' (in $T$-space) that lies
 beneath the combined heating and cooling
rates. 

\begin{figure}
\scalebox{0.45}{\includegraphics{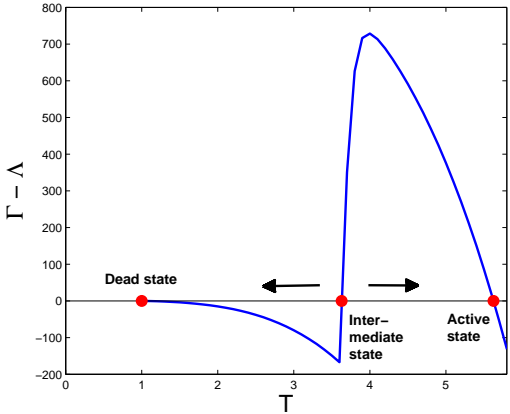}}
\begin{footnotesize}
 \caption{The combined heating and cooling rate as a function of
   temperature for the slaved model. Arrows indicate the direction in
   which gas will tend to move if subject to the thermal
   dynamics. We can easily see the two basins of attractions of
   $T_Q=1$ and $T_A$, with the intermediate state acting as the boundary
   between them. }
\end{footnotesize} 
\end{figure}

 The combined heating and cooling
$(\Gamma-\Lambda)$ 
 is plotted in Fig.~4. Quite generally there will be two opposing
 contributions to the area under this curve (the integral in
 \eqref{slavec}). 
There will be a positive contribution from the
 portion of the front lying near $T_A$, in the active state's basin of attraction.
 And there will be a negative contribution from the portion of the
 front near the lower quiet state, in its basin of attraction. Because
 the
 slaved model is only 1D, these two basins are
 easy to see. Gas that is a little hotter than the intermediate
 state is drawn to the upper turbulent state, while gas a little
 colder is drawn to the lower state: this is represented in
 Fig.~4 by the horizontal arrows. For the parameters chosen here
 the contribution from the active state's basin dominates
 and $c>0$.

Even if the integral in Eq.~\eqref{slavec} were intractable, we can at
 least say that the relative sizes of the two basins of attraction
 help determine the sign of $c$.
 For example, if the parameter $b/w$ is very
 small then $T_A =(1+w/b)^{1/4}$ will be large, and consequently the
 basin of attraction of the upper active state will also be large. We
 would then expect the integral \eqref{slavec} to be dominated by this region and
 for $c$ to be positive. On the other hand, if $b/w$ is continuously decreased
 so that $T_A$ approaches the intermediate state, then
 its basin of attraction will also decrease. At a second critical
 $b/w$ one would expect the sign of $c$ to change from positive to
 negative. 

These heuristic arguments can be worked through in some detail because
\eqref{slavec} is integrable
 once we treat
the heating $\Gamma$ as essentially a step function (cf. Section
4.1). We find
$$ \int_1^{T_A} (\Gamma-\Lambda)dT \approx \tfrac{1}{5}\,b\,T_A^5
-(b+w)T_A + w\,T_\text{MRI} + \tfrac{4}{5}\,b , $$
in which $T_A\approx (1+w/b)^{1/4}$. This equation permits us to
calculate the critical value of $b/w$ at which point $c=0$, as a function
of $T_\text{MRI}$. The equation corresponding to
$c=0$ is
\begin{equation}
(1+w/b)^{5/4} -\tfrac{5}{4}(w/b)\,T_\text{MRI} -1 =0,
\end{equation}
which must be solved numerically. However, because $b/w$ is small,
 a reasonable approximation
can be obtained by expanding the first bracketed term. This gives
$$ (b/w)_\text{cr} \approx (5\,T_\text{MRI}/4)^{-4} \approx
0.0025.$$
 It is easy to see that the system admits two families of MRI-fronts.
 When $(b/w) <  0.0025$ we have an outward
moving front, $c>0$. When  $0.0042 >(b/w) >  0.0025$ we have an inward
moving front, $c<0$. And when $(b/w)> 0.0042$, there is no front at
all (because there is no active state available, cf.\ Section 4.1). 
 Later we discuss how this parameter configuration $b/w$ depends on
 disk radius and, consequently, how an MRI-front changes as it
propagates through the disk. The fact that there is a value for $b/w$
for which $c=0$ means that there is likely a special
radius at which $c=0$, and hence an attracting radius towards which all
fronts migrate.

\begin{figure}
\scalebox{0.35}{\includegraphics{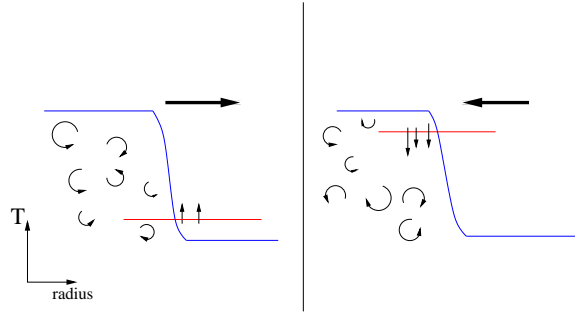}}
\begin{footnotesize}
 \caption{Simple cartoon of the temperature profile of two fronts. The
 red line overlaying them indicates the critical temperature
 separating the two basins of attraction of the hot/turbulent and
cold/quiescent states.}
\end{footnotesize} 
\end{figure}

Finally, we may obtain a better physical understanding
 on the sign of $c$ through the
following argument. 
In Fig.~5 we plot two hypothetical MRI-front profiles for
 the temperature $T$. Overlying these profiles is a red line which
 corresponds to the critical temperature $T_I$ separating the basins of attraction
 of the quiet and the active states. In the first picture the majority
 of the gas lies within the active zone's basin. Quiescent 
 fluid heavily perturbed
 by the neighbouring front will find itself more likely to be pushed
 over this critical threshold than fully turbulent fluid: the hot
 turbulent fluid will have to cool off much more than the cold fluid
 has to heat up. As this fluid heats up and becomes turbulent the
 front moves forward `gobbling' up this gas, and hence moves further
 into the quiescent state --- the large black arrow. In the second
 picture the situation is reversed and the MRI-front `retreats' into
 the turbulent gas.

\subsection{Numerical examples of MRI-fronts}

Having listed some analytic results in certain limits which help us understand the
nature of the dead-zone/active-zone interface, we now compute these
solutions explicitly from Eqs \eqref{frKK}-\eqref{bc2} using a
numerical scheme. 
 The problem is a 4th order nonlinear eigenvalue problem in one variable $\xi$, and we
expect the system to be quite stiff: the profiles will not change
much over most of the domain except near the transition front where
they will vary rapidly. As a consequence, we employ a relaxation
method,
 whereby an initial guess to the front profile and an initial guess for $c$
 are gradually `evolved', via Newton's method, to
 a numerical approximation of the correct solution (Press et al.~2007).

\subsubsection{MRI-front profiles}

We show results for which $b=0.001$ and $w$ varies. MRI-fronts travelling
either inward or outward are supported by the equations.
 Two examples are plotted in Figs 6 and 7. In both cases the
front joins the active and dead-zones smoothly over a characteristic
lengthscale. For most outward propagating fronts, the radial scale of the
transition region for both $K$ and $T$ is roughly $3 L_\text{T}$.
However, slower moving fronts (in either direction) or stationary
fronts exhibit
a temperature transition with a long `tail'
that extends tens of $L_\text{T}$ into the dead-zone.

\begin{figure}
\scalebox{0.5}{\includegraphics{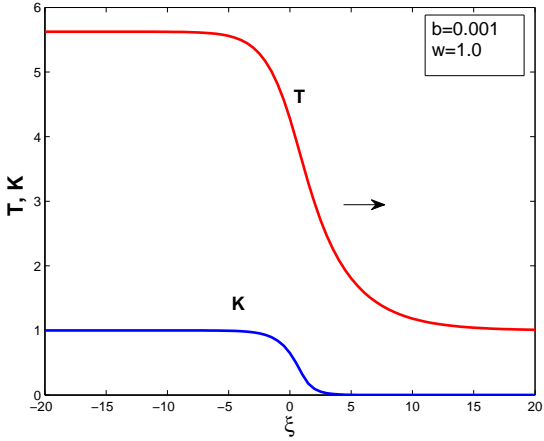}}
\begin{footnotesize}
 \caption{MRI-front profile with $P_r=1$,
   $b=0.001$, $w=1$. The front speed is $c=0.28 L_T\Omega$, so
  the front travels into the dead-zone, i.e. outward.}
\end{footnotesize} 
\end{figure}

\begin{figure}
\scalebox{0.5}{\includegraphics{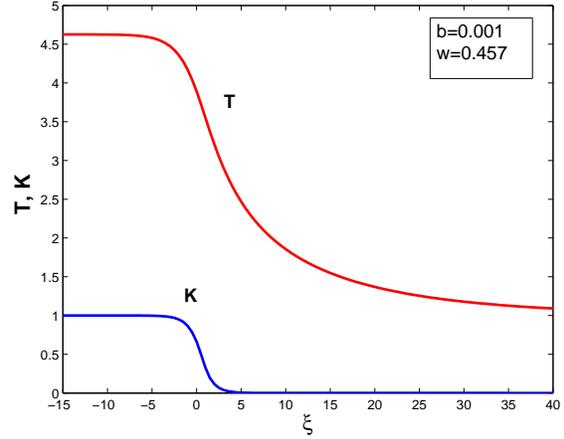}}
\begin{footnotesize}
 \caption{MRI-front profile with $P_r=1$,
   $b=0.001$, $w=0.457$. The front speed is almost zero $c= 9.3\times
   10^{-4}
    L_T\Omega$. Note the extended morphology of the
   temperature transition in comparison to $K$.}
\end{footnotesize} 
\end{figure}


\subsubsection{Front velocities}

Whether the front travels inward or outward its speed $c$ is bounded
according to $|c|< L_T\Omega < H\Omega$, and is thus
controlled by the largest turbulent eddies; naturally, the front cannot travel
faster than one diffusion length per orbit. 
The fastest fronts
are those with larger $w$, i.e. originating deeper
 in the active zone at small radii. These
may travel outward as fast as a disk scale height $H$ per orbit:
roughly 1 au in less than 10 years. However, $c$ will vary
as the front travels because of the dependence of the parameters ($w$
especially) on radius, as well as the physical scales $\Omega$ and
$H$.

\begin{figure}
\scalebox{0.5}{\includegraphics{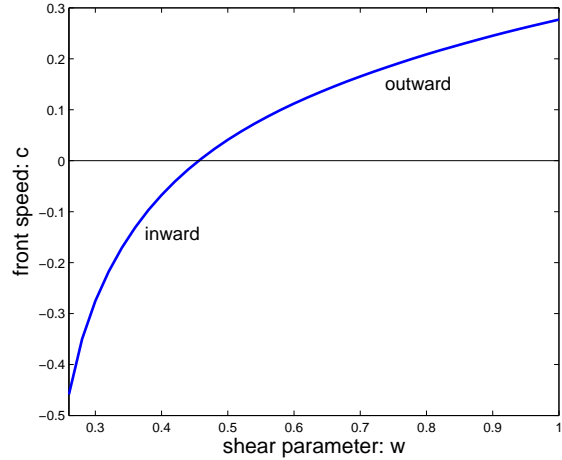}}
\begin{footnotesize}
 \caption{Front speed $c$ as a function of the parameter $w$. Other
   parameters are held constant, namely $b=0.001$. As $w$
   decreases so does $c$ until at a critical point $c=0$, after which
   $c$ is negative and the front propagates inward. The critical $w$
   is approximately $w=0.4563$.}
\end{footnotesize} 
\end{figure}

In Fig.~8, we show how the front speed $c$ changes as we vary the 
shear parameter $w$, keeping $b$ fixed. As $w$
   decreases so does $c$ until at a critical point $c=0$ after which
   $c$ is negative and the front propagates inward. At a second value
   of $w$ the front ceases to exist: the upper turbulent state
   vanishes and the gas is no longer bistable. This behaviour adheres to our
   expectations given the simple slaved system of Section 5.2. There
   the
   critical value of $b/w$ is 0.0025, at which point $c=0$. The more
   general system here yields a value closer to 0.0022, which is in fair
   agreement.

 We
   associate smaller $w$ values with larger radii (less shear) and
   larger $w$ with smaller radii (greater shear). It is expected then
   that fronts originating at larger radii propagate inwards and
   fronts coming from smaller radii propagate outwards. In between
   there is a critical radius $R_\text{att}$ at which point $c=0$
   and the front stays
   put. It follows that \emph{all} MRI-fronts are attracted to
   this special radius. And this
radius by necessity lies within the bistable region:
$\Rr<R_\text{att}<\Rt$.
 Consider a front
that begins near the inner edge of the bistable zone $\Rr$. It will
travel outward at some initial speed, but as it approaches
$R_\text{att}$ it will decelerate; eventually it will slow to a crawl
and come effectively to a halt near this radius. On the other hand, a
front that begins near the outer boundary of the bistable zone $\Rt$ will
propagate inwards until it too approaches $R_\text{att}$, at which
point it slows down and eventually stops.

\section{Travelling fronts in global disk models}

In this section the physics explored in the previous sections are
tested in simulations of the full equations 
\eqref{finK}--\eqref{finT} in a simple global disk model. 
 As we have argued, an MRI-front usually does
not linger in
one place; it will by its nature move away to radii where the properties of
the gas are going to be different to where it started, and
consequently where its profile and speed will change. How this works
in detail can only be captured by a calculation that take into account
the radial structure of the disk.
 
The
geometric cylindrical terms are reinstated in \eqref{finK}--\eqref{finT}
and the disk's radial
 structure is encoded
 through the dependence of the cooling and shear 
parameters $b$ and $w$
 on radius $R$. By setting $w \propto \Omega$ and recognising that the
 cooling time decreases with radius, we suggest the following model
\begin{align} \label{global}
w = w_0\left(\frac{R}{\Rt}\right)^{-3/2}, 
\qquad b = b_0 \left(\frac{R}{\Rt} \right)^{p},
\end{align}
where the constants $w_0$ and $b_0$ correspond to the critical values
for which there are no active states, and at which point $R=\Rt$. 
Recall from Section 3, that this occurs when $b_0/w_0 = 0.0042$. The
power $p \geq 0$ we are free to set, but given the uncertainties in the
opacity in the inner disk we set $p=0$ in most applications.
In
addition, we should let $T_\text{eq}$ also be a function of $R$ because of
heating from the central star. This aspect of the problem is neglected
mainly for ease, and also because appreciable change in $T_\text{eq}$
should be limited to near the inner disk edge. 

The equations \eqref{finK}--\eqref{finT} are simulated with a
finite-difference scheme and explicit first order time-stepping over
a radial domain extending from 0.1 au to 5 au. Space units are chosen
so that $\Rt=2$ au and $L_T= 0.1$ au. Time is scaled by the orbital
period at $\Rt$. Various initial conditions were
trialed, including simple tanh profiles. But all yielded the same long
term equilibrium.

\begin{figure}
\scalebox{0.55}{\includegraphics{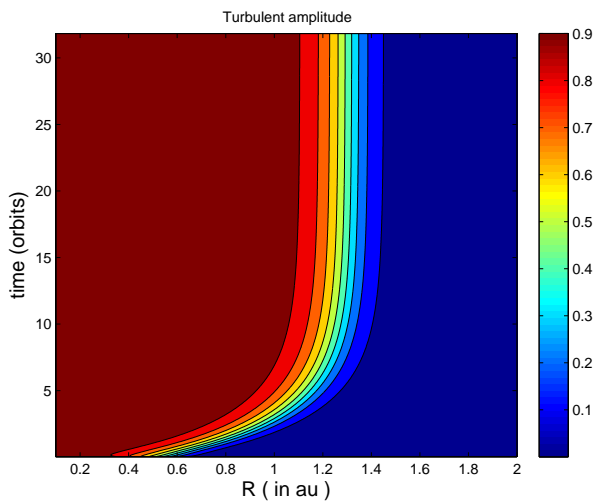}}
\begin{footnotesize}
 \caption{Space-time diagram of the evolution of the turbulent
   amplitude $K$ in a
   global model of a protoplanetary disk. The front propagates from
   near the inner edge and settles near the radius $R_\text{att}$,
   which is some 65 \% of $\Rt=2$ au.}
\end{footnotesize} 
\end{figure}

\begin{figure}
\scalebox{0.55}{\includegraphics{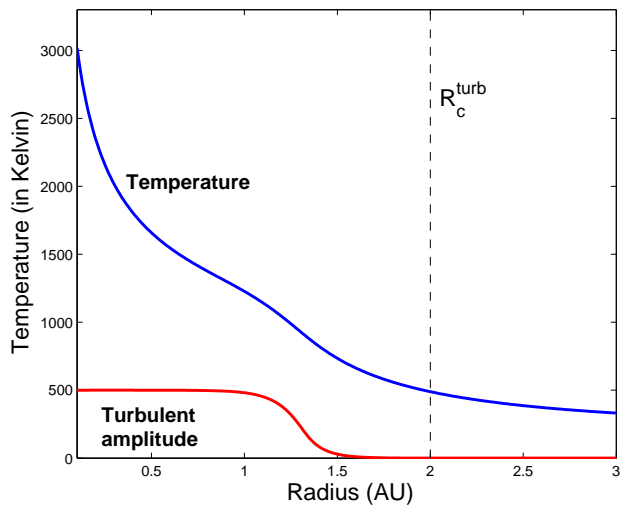}}
\begin{footnotesize}
 \caption{Long time equilibrium of the temperature field in a global
   model of a protoplanetary disk. Superimposed on the figure is the
   turbulent amplitude rescaled by a factor 500 so as to be visible.}
\end{footnotesize} 
\end{figure}

In Fig.~9 we plot a typical 
space-time diagram of the turbulent amplitude $K$. The
front is released at a small radius, near the inner edge, from where
it 
propagates outward rapidly, covering 0.6 au in less than 5 orbits. It
then slows down as it approaches the critical radius $R_\text{att}$
and eventually comes effectively to a halt. In Fig.~10 we plot a
snapshot of the
long time temperature profile, obtained after 200 orbits. The front is
completely stationary at this point. Note that, unlike the turbulent
amplitude, the temperature $T$ varies with radius even far from the
location of the front. This is because the turbulent equilibrium state
$T_A$ is a function of radius through $w$. Thus at smaller radius the turbulent heating is 
greater, as is $T_A$ (cf.\ \eqref{EqmApprox}). Note that the model appears to overestimate
$T$ at these small radii in comparison with alpha models (see e.g.\ Terquem 2008).
 Superimposed upon the figure is the turbulent amplitude,
scaled appropriately. 

In these figures, and more generally, the final location of the front
$R_\text{att}$ 
is roughly 65\% of $\Rt$. This could be pre-empted from
Eq.~\eqref{global}, once we recognise that
 $b/w = 0.0022$ at this radius. Then
we have the estimate
\begin{equation}
\left(\frac{R_\text{att}}{\Rt}\right) = \left(\frac{0.0022}{0.0042}\right)^{2/(3+2p)},
\end{equation}
which gives $\approx 0.65$ when $p=0$. For $p>0$ the attracting radius
is greater.

Finally,
 the front profiles we have simulated appear to be \emph{stable}
even when they slow to a
static equilibrium. The simulations show no evidence of
secondary instability and a subsequent
 dissolution into more disordered time-varying dynamics as
 witnessed in, for example, flame fronts
 (Zeldovich and McNeill~1985). Neither is there any sign of the
 instability uncovered by
 Wunsch et al.\ (2005, 2006). If such instabilities exist in real
 protoplanetary disks, our model is too simple
 to capture them, omitting, for example, accretion. 
But there is also the question of consistency.
 Our model, and the related alpha disk models, have been
derived via averaging over small-scale
 fluctuations, and have thus assumed that these fluctuations
 are well-behaved and bounded.
  The issue of front instability and
 more complicated time-dependent behaviour must be settled by direct
 numerical simulations of the MRI itself.

\section{Summary and discussion }

We now summarise the contents of the paper
 and bring out its most important points.
 In Section 2, we made the case that in most
protostellar disk systems, midplane gas at inner radii (less than
about 1 au) finds itself bistable: it can be either quiescent or
MRI-turbulent.  As a consequence, the location of the
dead/active zone boundary may be difficult to pin down and, moreover,
the boundary profile itself not particularly defined. Estimates
on the critical radii which bound this gas are not well constrained;
but we expect the inner radius $\Rr$ to be (much) less than 0.1 au, and the
outer radius $\Rt$ to be roughly 1 au, with its exact value dependent
on stellar mass and accretion rate. 

In order to understand this region, a reduced model was introduced
(Section 3) which describes
 the mutual coupling of the turbulent and thermal dynamics of
the gas in the bistable region. The turbulence is modelled in a
mean-field fashion, which assumes that its fluctuations are
well behaved: large amplitude intermittent behaviour (which might be
quite important) is smoothed out. 
The governing equations take
the form of two reaction-diffusion equations for temperature and
turbulent amplitude. After a suitable rescaling, the remaining free
parameters have relatively transparent meanings: $b$ represents the
local gas opacity, and $w$ represents the local shear.

In Section 4, we set this model to work computing homogeneous
equlibrium states, in particular showing that in the bistable zone of
a disk there should be three equilibria: the dead-zone state and the
active turbulent state (both stable as expected) in addition to an intermediate
state that is unstable and which polices the boundary between these
two outcomes. 

In Section 5 we compute what we call `MRI-fronts', which are
travelling transitions connecting the active zone to the dead zone. We
can prove a number of properties about these fronts. Of greatest
importance is the sign of the front speed: we show
that fronts beginning at the inner edge of the bistable zone
propagate outward, while fronts beginning at the outer edge of the same
zone travel inwards. There then exists a critical radius
$R_\text{att}$ 
at which
point the MRI-front velocities are precisely zero and it is towards
this radius that all fronts must go and finally anchor themselves.
This is the true effective boundary between active and dead zones.
Numerical simulations of
the equations in a global model of a protoplanetary disk are given
 in Section 6 and these confirm this expectation.
They also show that the front solutions are stable to secondary
perturbations, at least within the confines of our model assumptions.

So where does this leave us? Our simple system predicts 
that the interface between MRI-active and inactive regions is in fact a robust
and coherent feature that relaxes into a steady
profile. Large deviations from this steady equilibrium will
decay and the MRI-front will ultimately return to its halting radius
$R_\text{att}$. This critical radius can be somewhat less than
existing estimates, but the idea that there exists a
well-defined boundary still applies. 
The picture that we have developed is certainly favourable
 to planet
formation scenarios that rely on a stable and quasi-static
dead/active zone interface. If the interface is as robust as suggested
by the model,
its associated pressure inhomogeneity may
be equally robust and long-lived: precisely what is required for the
accumulation and eventual agglomeration of solid material.

However, the model has neglected (in fact, `averaged away') the highly
intermittent turbulent fluctuations that we expect such an MRI-front
to suffer. The smooth profiles that we have generated (Figs 6 and
7) represent an average;
in reality, an MRI
front will most likely be a disordered structure subject to
potentially large amplitude perturbations.  
Moreover, it is not quite clear what lengthscales 
this disorder encompasses --
it will probably be on a scales of order the transition scale itself -- some
fraction of $H$. If
the transition region is indeed volatile and fairly broad then it may
inhibit the development of the pressure inhomogeneity required by
planet formation theories. Moreover, additional physics
(accretion, for example) and
realistic vertical structure may
stimulate instability and associated secondary fluctuations.
A sensible way to check some of these effects
is to better describe the turbulent dynamics and
 move on from the diffusive approximation.  
MHD simulations of the MRI with a temperature
dependent resistivity are needed. These might be undertaken in
 cylindrical geometry, which
 is sufficient to capture the key global effects (with shear rate
 declining with radius), while allowing acceptable resolution, given modern computational
constraints. Such simulations are planned and will be presented in
later work.

\section*{Acknowledgments}
The authors would like to thank the anonymous reviewer for a thorough
set of comments that helped improve the presentation of the paper.
This work has been supported by a grant from the Conseil R\'egional
de l'Ile de France and STFC grant ST/G002584/1.
This work greatly benefited from the comments and criticisms of Pat
Diamond (during ISIMA 2010), Pierre Lesaffre, Tobias Heinemann, and in particular Sebastien
Fromang who read through a preliminary version of the manuscript.

\appendix

\section{Derivation of the mean-field model}

Though the mean-field model used throughout the
main paper can be motivated by physical heuristic arguments, its
general lineaments can
formally be derived through a separation of scales analysis. In this
appendix we briefly sketch out the procedure.

\subsection{Governing equations}

We begin with the full equations of resistive and viscous MHD 
\begin{align}
&D_t \rho = -\rho\,\nabla\cdot \u, \label{ge1}\\
&D_t\u = -\nabla \Phi -\frac{1}{\rho}\left(P+\frac{B^2}{8\pi}\right)
+\frac{1}{4\pi\rho}\B\cdot\nabla\B + \frac{1}{\rho}\nabla\cdot\mathbf{\Pi}, \label{ge2}\\
&D_t\B = \B\cdot\nabla\u - \B\,\nabla\cdot\u + \eta
\nabla^2\B, \label{ge3}\\
&E\,D_t S = \rho( \psi_\text{dis}+ \psi_\text{ext}) - \nabla\cdot\FF ,\label{ge4}
\end{align}
where $\rho$ is volumetric mass density, $\u$ is velocity, $\Phi$ is the
gravitational potential of the star, $P$ is pressure, $\B$ is magnetic field,
$\mathbf{\Pi}$ is the (molecular) viscous stress tensor, $\eta$ is the
resistivity, $E$ is the internal energy density,
 $S=\ln P\rho^{-\gamma}$ is the entropy function and $\gamma$
the ratio of specific heats, $\rho\psi_\text{dis}$ comprises viscous and
resistive heating, and $\rho\psi_\text{ext}$ represents external heating
from the star. Finally, the radiative flux is denoted by $\mathbf{F}$.
We also have denoted the total derivative by $D_t= \d_t +
\u\cdot\nabla$. In writing the above we have assumed an ideal gas equation
\begin{equation}
P= \frac{\mathcal{R}}{\mu}\,\rho\,T
\end{equation}
where $\mathcal{R}$ is the gas constant and $\mu$ is mean molecular
weight. It follows that $E= P/(\gamma-1)$. We set
 the resistivity to be a function of temperature
$\eta=\eta(T)$ via an appropriate form of Saha's equation (see Stone et al.~2000 and 
Balbus 2011). This means
that we are considering thermal ionisation explicitly. We neglect
non-thermal ionisation.

\subsection{Fluctuations}

We assume that equations \eqref{ge1}-\eqref{ge4} admit a laminar equilibrium state
characterised by (near) Keplerian rotation:
$$ \u = \u_0(R) = R\Omega(R)\,\ePhi $$
where $R$ is cylindrical radius (the tiny radial drift arising from
molecular viscous torques we omit). There will also be some radial and
vertical structure in the variables $\rho$ and $P$,
 which won't be listed. Finally, we assume that there is a weak magnetic field
$\B_0$ that does not influence the equilibrium balances significantly
but which plays the defining role in the disk's stability by
catalysing the MRI. 

In a given region of this disk the MRI will occur if the local
resistivity is sufficiently low. We next assume that in such an
unstable region turbulence will ensue, giving rise to both magnetic
and velocity fluctuations which we denote by $\v$ and $\b$. Because
MRI turbulence is mainly subsonic, associated
density fluctuations are considered to be small and are neglected.

The
velocity
 fluctuations are small compared to the Keplerian motion
$ |\v| \ll |R\Omega| $
and inhabit a well-defined range of lengthscales, the upper bound of which
we denote by $l$, understood as either the main
 energy injection scale of the MRI,
or the scale of the largest turbulent eddies. We then
posit the following hierarchy
$ l \ll H \ll R,$
where $H$ is the disk semithickness. The first scaling is marginally
true at best, and in fact we can only probably say $l<H$, but the
stronger form is necessary for the mathematical argument to go
through.
This motivates the introduction
of the small parameter 
$ \epsilon = l/H.$

\subsection{Fluctuation equations}

The
magnitude of the fluctuations is defined through
$$ k = \tfrac{1}{2}[ \rho \v^2 + \b^2/(8\pi) ] $$
which we term the `turbulent amplitude' or `intensity'. 
One can derive an evolution equation for $k$ from Eqs
\eqref{ge1}-\eqref{ge4}. This can take multiple forms (see for example
Balbus and Hawley 1998). We use the following:
\begin{equation} \label{unavKeq}
\d_t k + \nabla\cdot\left(\v \,k\right) = q,
\end{equation}
where the source term $q$ is rather ungainly. It can be written as
\begin{align}
q &= \v\cdot(\nabla\cdot\mathbf{\Pi}) - \v\cdot\nabla P -
\nabla\cdot\left(\u_0\,k\right) \notag \\
&\hskip2cm  +\frac{1}{4\pi}\left[\eta \b\cdot\nabla^2\b +
  \v\cdot(\v\cdot\nabla \u_0)\right. \notag \\ 
&  \left.+ \b\cdot(\b\cdot\nabla \u_0)
  +(\b+\B_0)\cdot\nabla(\b\cdot \v) -\v\cdot\nabla(\B_0\cdot\b) \right].
\end{align}
This $q$ term is obviously an exceedingly complicated function of $k$. It
both controls the emergence of the MRI and its nonlinear saturation
through turbulence. Note that it also depends explicitly on temperature
through $\eta$, and to a lesser extent $P$.

In addition, we can derive an equation for fluctuations in internal
energy or temperature. It is more convenient, however, to deal with
the total temperature $T$. It is
\begin{align}\notag
\d_t T & + \nabla\cdot\left(\v\,T\right) = 
-\nabla\cdot(\u_0\,T) +(2-\gamma)T\nabla\cdot\u \\
& \hskip0.7cm+ \frac{(\gamma-1)}{\mathcal{R}/\mu}\left[
\psi_\text{dis}+\psi_\text{ext} - \frac{1}{\rho}\nabla\cdot\mathbf{F} \right].
 \label{unavEeq}
\end{align}
In both Eqs \eqref{unavKeq} and \eqref{unavEeq} the left hand 
flux term comprises only the advection from the turbulent velocity
field.

\subsection{Small scales and averages}

We wish to examine the dynamics on scales longer than the
fluctuations, though shorter than the radial scale of the entire
disk. Basically, we want to look at stuff happening in the
bistable region which has a radial size on an intermediate scale $\lesssim 1$ au.
We hence introduce a spatial average over (a) azimuth, (b) the vertical extent
of the disk, and (c) over the short radial scales of order $l$.
This procedure effectively smooths out
turbulent fluctuations and reduces them eventually to a turbulent flux, 
a violation of some of the complexities of the flow, but a procedure that
furnishes relatively simple evolution equations for the two quantities
of interest $k$ and $T$. Note that, after averaging, these 
quantities depend only
on $t$ and $R$, where the latter coordinate describes 
radial scales longer than
the turbulent eddy size $l$.

In order to define proper averages over the turbulent fluctuation scale $l$
at a
given point in the disk,
a short radial coordinate is necessary. This we denote by $x$ and
define through
$$ R = R_0 + \epsilon\, x $$
where $R_0$ is the radius in the disk in which we are interested. We
may now regard any fluctuating quantity (such as $\v$ and $\b$) as
depending on both $x$ and $R_0$ (in addition to $\phi$, $Z$, and
$t$). The $R$ derivatives of these functions must then be replaced by
$R_0$ and $x$ derivatives:
$$ \frac{\d}{\d R} \to \frac{\d}{\d R_0} +
\frac{1}{\epsilon}\,\frac{\d}{\d x}.  $$

A multiple spatial average is introduced. For a
quantity $f=f(x,R_0,\phi,Z,t)$ we have
\begin{align}
\langle f\rangle =
\left(\frac{1}{2^3\,\pi\,H\,L}\right)
\int_0^{2\pi} d\phi\, \int_{-H}^{H} dZ\,\int_{-L}^{L} dx\,f\,,
\end{align}
where $L$ is an intermediate scale of order $H$ upon which we wish to
investigate the thermal/turbulent dynamics. It follows then that the
averaged quantity $\langle f \rangle$ depends on only $R_0$ and $t$. 

\subsection{Mean field equations}

We first attack the evolution equation for $k$,
i.e. Eq.\eqref{unavKeq}, using the average defined in the previous
subsection. The most problematic term is the flux 
$\langle\nabla\cdot(\v\,k)\rangle$. We first note the
following
\begin{align*}
 &\int_0^{2\pi} \d_{\phi}(v_\phi\,k) d\phi = [v_\phi\, k]^{2\pi}_0=0 \\
 &\int_{-H}^H \d_z( v_z\, m) dz = [v_z\, m]^H_{-H} =0,
\end{align*}
where the first equality holds due to periodicity in $\phi$, and the
second holds providing that velocity fluctuations go to zero at the
disk's upper and lower surfaces. This leaves the $R$ derivative term,
$$ \int_{-L}^L \d_R (R\,v_R\,k) dx = \frac{\d}{\d R_0} \int_{-L}^L R v_R k dx + 
\epsilon^{-1} [R v_R\, k]^L_{-L} .$$
Recognising that $l\ll L$, the velocity fluctuations $v_R$ are
uncorrelated at $x=\pm L$. Also $L\ll R$ so that $R\approx R_0$. This
means that the second `surface' term is merely a source of `white
noise' to leading order and thus contributes no systematic
bias over intermediate and long times. Formally it
can be averaged out by an additional temporal average over short
times. Here we just let it equal zero.

In summary,
\begin{equation}
\langle \nabla\cdot(\v k) \rangle \approx \frac{1}{R_0}\frac{\d}{\d
  R_0}
\left(R_0 \langle v_R \, k \rangle  \right),
\end{equation}
and so the averaged turbulent amplitude equation becomes
\begin{equation} \label{avKeq}
\frac{\d}{\d t} \langle k \rangle+ \frac{1}{R_0}\frac{\d}{\d
  R_0}
\left(R_0 \langle v_R \, k \rangle  \right) = Q, 
\end{equation}
where $Q=\langle q \rangle$. This is an evolution equation for the averaged turbulent amplitude
$\langle k \rangle$, which we now relabel $K$.

An analogous procedure gives an equation for the average temperature,
\begin{equation} \label{avEeq}
\frac{\d}{\d t}\langle T \rangle + \frac{1}{R_0}\frac{\d}{\d
  R_0}
\left[R_0\,\left( \langle v_R \, T \rangle +  \Theta\right) \right] =
 \Gamma -\Lambda,
\end{equation}
where the radial radiative flux $\Theta$ is given by
\begin{equation}
\Theta= \frac{(\gamma-1)}{\langle
    \rho\rangle\mathcal{R}/\mu}
\langle F_R\rangle,
\end{equation}
the `turbulent heating' function is
\begin{equation} \label{psi}
 \Gamma = \frac{(\gamma-1)}{\mathcal{R}/\mu}\langle\psi_\text{dis} \rangle -(2-\gamma)\langle T\nabla\cdot \v \rangle 
\end{equation}
 and
\begin{equation}\label{lambda}
 \Lambda = \frac{(\gamma-1)}{\mathcal{R}/\mu}\left[ \langle \psi_\text{ext} \rangle + 
\frac{1}{2^3 H L \pi}\int_{0}^{2\pi}\int_{-L}^L
[F_z]^{H}_{-H}/\langle\rho\rangle\,d\phi dx\right], 
\end{equation}
is a `cooling' function comprising the external source and radiative
losses from the disk surfaces. Unfortunately, in order, to get the
radiative flux and cooling term in this form,
it is necessary to treat the $1/\rho$ factor of
$\nabla\cdot\mathbf{F}$ in \eqref{unavEeq} as $\langle 1/\rho\rangle$,
i.e. to use a `pre-averaged' value for the density, instead of
properly taking account of its variation in the full integral. It is
doubtful, however, that this introduces any appreciable change to the
cooling function $\Lambda$ itself. Henceforth we suppress the angle
brackets on $\langle
T\rangle$. 

In order to make any progress with Equations \eqref{avKeq} and
\eqref{avEeq} we need to know how the functions $Q$, $\Gamma$, and $\Lambda$,
and the fluxes $\Theta$, $\langle v_R\, T \rangle$, and 
$\langle v_R\, k \rangle$ behave as
 $K$ and $ T $ vary. Until we know this,
 we cannot proceed. This is the closure problem. 
 We prescribe the form of these functions
 from simple but physically plausible assumptions. In Section 3.1 some
 of these choices are discussed, we expand on this in the following
 subsections.

\subsection{Fluxes: the diffusion approximation}

 Because the disk is optically thick at the
radii in which we are interested ($\sim 1$ au) we are permitted to
employ the diffusion approximation for the radiative flux of
energy. That is to say that energy radiates preferentially down a
temperature gradient. Consequently,
\begin{equation}
 \Theta \approx - D_T^{\text{rad}}\, \frac{\d T}{\d R_0} ,
\end{equation}
where the diffusion coefficient is a function of temperature.

The turbulent fluxes are naturally more difficult. We  employ the
simplest turbulent closure in this instance --- that of the eddy
viscosity model --- in which turbulence simply transports gas
quantities down gradients in those quantities. If both $T$ and $k$
were proper passive scalars then this assumption could be made more
strongly using an additional multiscale method (see Frisch 1989), or a Lagrangian
type argument (see Balbus 2011). Unfortunately they are \emph{active}
scalars and interact with and vary along the eddies that convect them.
We believe however that this model suffices to get a handle on the
 basic qualitative behaviour in which we're interested.
We take the following:
\begin{align}
\langle v_R\, T \rangle \approx - D_T^\text{turb}\,\frac{\d T}{\d
  R_0} , \quad
 \langle v_R\, k \rangle \approx - D_K\,\frac{\d K}{\d R_0},
\end{align}
where the diffusion coefficents $D_T^\text{turb}$ and $D_K$ are
functions of the local (averaged) turbulent magnitude $K$.
 Theoretical issues with this approach are raised and
discussed in Section 3.5. 

Lastly we examine
the relative magnitudes of turbulent and radiative diffusion of heat. 
Though MRI turbulence will certainly mix
heat and will spread its turbulent motions, these transport processes
have not been studied in any depth (see, however Hirose and Turner
2011 for an examination of the optically thin upper layers of a disk).
 Numerical work has focused mainly
on the transport of angular momentum and small
particles. For instance, Carballido et al.~(2005) measure the diffusion
of a passive scalar in MRI turbulence simulations and find the
diffusion coefficient to be $\sim 10^{-2} H^2\Omega$. This quantity,
however, depends on the magnetic field geometry and strength and can
take larger values (Johansen et al.~2006).
 Adopting a minimum mass solar
nebula, we estimate the
turbulent diffusion coefficients to be
$$ D_K \sim D_T^\text{turb} \sim \alpha_\text{th} H^2 \Omega \sim \alpha_\text{th} 10^{16}
\text{cm}^2\,\text{s}^{-1}, $$
at 1 au, where $\alpha_\text{th}$ is the `turbulent alpha' for thermal
transport $< 10^{-2}$.
These coefficients drop to zero, of course, in the quiescent state.

Radiative diffusion, on the other hand, is easier to constrain. We have 
\begin{align*}
 D_T^\text{rad} &\sim \frac{16(\gamma-1) \sigma_{\text{SB}}\, T^4}{3 \kappa
  \rho P}\\ 
&\sim
2.39\times 10^{12}\,\left(\frac{T}{100\,\text{K}}\right)^3\left(\frac{\text{cm}^2\,\text{s}^{-1}}{\kappa}\right)
\text{cm}^2\,\text{s}^{-1},
\end{align*}
in which we have set $\rho\sim 10^{-9}$
g cm$^{-3}$. The Stefan-Boltzmann constant is $\sigma_{\text{SB}}$.
It follows that in a hot turbulent state ($\sim 1000$ K),
radiative diffusion is comparable to turbulent
diffusion. In the quiescent
state, however, radiative diffusion dominates,
as then $D_T^\text{turb}$ goes to zero. Note that beyond the
dust sublimation temperature $\gtrsim 1600$ K the opacity will vary
steeply and, as a consequence, radiative diffusion will certainly
dominate. However, it is expected that turbulent heating sustains such
temperatures at very small radii ($\lesssim 0.1$ au; see Terquem 2008,
for example). In the present work we do
not attempt to model the complexities of the transition from one
regime to the other and instead treat the diffusion coefficients as constants.

\subsection{Reaction terms}

We address $\Lambda$ first, which determines radiative equilibrium:
 the temperature balance in
the absence of turbulent dissipation. Radiative losses on the disk surfaces can
be approximated by
$$ \frac{1}{2^3 H L \pi}\int_{0}^{2\pi}\int_{-L}^L
[F_z]^{H}_{-H}\,d\phi dx \approx 2\sigma_\text{SB} T^4_\text{surf}, $$
where $T_\text{surf}$ is a radially and azimuthally averaged effective surface
temperature.
We relate $T_\text{surf}$ to the midplane temperature $T_c$ through
$ T_\text{surf}^4 = 4\,T_c^4/(3\, \tau_c) $, as is customary,
where $\tau_c$ is half the total optical thickness. The 
central temperature $T_c$ is then assumed to be proportional to
the vertically averaged
temperature $T$. Putting this into
\eqref{lambda} and rearranging gives
\begin{equation} \label{cool}
\Lambda = b\,(T^4 - T^4_\text{eq} ),
\end{equation}
where $T_\text{eq}$ is the temperature at radiative equilibrium and $b$ is
some free parameter associated with the opacity of the gas. It
controls, in basic terms, the speed at which radiative equilibrium is
enforced. 

The MRI terms are represented by the Landau operator $Q=sK- a K^2$,
where $s$ is the linear growth rate of the MRI. 
It is assumed that temperature variations only
intrude on the model via this linear forcing term. Smaller
temperatures mean greater resistivities, and hence lower (or zero) 
growth rates. From consideration of the linear dispersion relation of
the resistive MRI,  Lesaffre et
al.~(2009) accounted for
this via
$s = s_0[1- \overline{\eta}(T)]$
where $s_0$ is the ideal MHD growth rate, $\overline{\eta}$ is the
non-dimensionalised resistivity.
 From the Saha equation
a plausible form for $\eta$ might be
\begin{equation} \label{eta11}
\eta \propto \,T^{-1/4}\,\text{exp}(T^*/T),
\end{equation}
for $T^*$ a constant. However, the extreme gradients, and
massive decay rates that can ensue, mean that the prescription
\eqref{eta11} has terrible numerical properties. 
 A more convenient model for $\eta$ is
\begin{equation}\label{etaa}
\eta \propto 1-\text{tanh}[6(T-T_\text{MRI})/T_\text{eq}],
\end{equation}
at least in the context of our reduced system.
 The tanh function captures the `switch'-like
behaviour of the Saha equation at $T_\text{MRI}$ but does not exhibit
a negative and divergent $s$ when $T$ is low. The choice of
coefficient $6$ gives the best comparison with \eqref{eta11} in the
regimes of interest. For instance, the maximum relative error that emerges in $s$ is
some 6\%.

\section{Linear stability of homogeneous states}

 We investigate purely homogeneous perturbations
 so that
$$ K= K_0 + K_1(t), \qquad T= T_0 + T_1(t) $$
where $K_0$ and $T_0$ is the equilibrium in question and $K_1$ and
$T_1$ are small perturbations. The linearised equations of \eqref{finK}
and \eqref{finT} are
\begin{align}
\sigma\,K_1 &= s_0\,K_1 - s'_0\,K_0\,T_1- 2\,K_0\,K_1, \\
\sigma\,T_1 &= w\,K_1 - 4b\,T_0^3\,T_1, 
\end{align}
in which we have decomposed $T_1$ and $K_1$ into Fourier modes
$\propto e^{\sigma t}$, where $\sigma$ is a linear growth rate (not to
be confused with the MRI growth rate $s$). Also
$s_0$ and $s_0'$ correspond to $s$ and $ds/dT$ evaluated
at $T=T_0$. The following second order dispersion relation ensues:
\begin{equation}
\sigma^2 +(K_0^2+ 4 b\,T_0^3)\sigma + (4 b\,T_0^3 -w s_0')K_0 = 0.
\end{equation}
Linear stability of the state $(K_0,\,T_0)$ is assured if both 
the coefficient of $\sigma$
and the last term are positive. The first condition is always
satisfied, which leaves us with the main stability criterion:
\begin{equation}
 4\,b\, T_0^3 - w\,s_0' >0.
\end{equation}
This can be rewritten into a more illuminating form by using the
cooling function $\Lambda$ and the
heating function at equilibrium
$$\Gamma(T)= w K = w\,s.$$
Then the stability criterion becomes
\begin{equation}
\frac{d\Lambda}{d T} > \frac{d\Gamma}{d T},
\end{equation}
which says that the cooling rate must outstrip
the heating rate if the state is to be stable, in
 accord with the intuitive
argument given in Section 4.2.

\end{document}